\documentclass[onecolumn,showpacs,pra,aps]{revtex4}
\usepackage{dcolumn}
\usepackage{bm}
\usepackage{graphicx}
\usepackage{mathrsfs}
\begin{document}
\title{Spatial Resolution Enhancement in Quantum Imaging beyond the Diffraction Limit Using Entangled Photon-Number State}
\author{Jianming Wen,\footnote{Electronic address: jianm1@umbc.edu} Morton H. Rubin, and Yanhua Shih}
\affiliation{Physics Department, University of Maryland, Baltimore County, Baltimore, Maryland 21250, USA}

\date{\today}

\begin{abstract}
In this paper we study the resolution of images illuminated by sources composed of $N+1$ photons in which one non-degenerate photon is entangled with $N$ degenerate photons. The $N$ degenerate photons illuminate an object and are collected by an $N$ photon detector. The signal from the $N$ photon detector is measured in coincidence with the non-degenerate photon giving rise to a ghost image. We discuss the case of three photons in various configurations and generalize to $N+1$. Using the Rayleigh criterion, we find that the system may give an improvement in resolution by a factor of $N$ compared to using a classical source. For the case that the $N$-photon number detector is a point detector, a coherent image is obtained. If the $N$-photon detector is a bucket detector, the image is incoherent. The visibility of the image in both cases is $1$. In the opposite case in which the non-degenerate photon is scattered by the object, then, using an $N$-photon point detector may reduce the Airy disk by a factor of $N$.
\end{abstract}


\maketitle

\section{Introduction}
Diffraction puts a limit on the the resolution of optical devices. According to the Rayleigh criterion \cite{rayleigh,goodman}, the ability to resolve two point sources is limited by the wavelength of the light. The Rayleigh or diffraction limit is not an absolute limit and proposals to exceed it have been known for a long time
\cite{goodman}. Recently, new proposals to improve resolution beyond the Rayleigh limit have been made based on the use of entangled sources and new measurement techniques. Improving the resolving power of optical systems beyond the diffraction limit not only is of interest to the fundamental research, but also holds promise applications in remote sensing and quantum sensors.

Classical imaging can be thought of as a single photon process in the sense that the light detected is composed of photons each of which illuminates the object, consequently, the image can be constructed one photon at time. What we mean by referring to this as classical is that the source of the light may be described by a density matrix with a
positive P-function \cite{pos1,pos2}. In this sense the Rayleigh limit may be thought of as a single photon limit. Recall that ideal imaging is a process in which there is a point-to-point mapping of the object to a unique image plane. Diffraction causes each point of the object to be mapped onto a disk, the Airy disk, in the image plane.

One of the new approaches to improving resolution is based on using non-classical light sources. Quantum ghost imaging
\cite{todd,streklov,rubin,imaging,shih,milenaL} is a process that uses two-photon entanglement. The unique features of this process are that entanglement allows only one photon to illuminate the object while the second photon does not. All the photons that illuminates the object are detected in a single (bucket) detector that does not resolve the image. The point detectors that detect the second photon must lie in a specific plane. This plane is called the image plane
although there is no image in that plane; the image is formed in the correlation measurement of entangled photons. The image is constructed one pair at a time. The resolution of this system has recently been discussed \cite{rubin2008,milena}. Losses in this system affect the counting rate but not the quality of the image.

A second approach using non-classical source is based on entangled photon-number states \cite{dowling}, e.g., N00N state. When the number of entangled photons exceeds two there are many possible imaging schemes that can be envisioned and so the analysis of these cases is still being carried out. This interferometric approach achieves a sub-wavelength spatial resolution by a factor $N$ and requires an $N$-photon absorption process. Another quantum source used to study imaging is to generate squeezed states \cite{squeezed}. The image can be reconstructed through the homodyne detection \cite{homodyne}. However, both of these techniques are severely limited by the loss of photons.

A second class of approaches to improving resolution uses classical light sources. One method uses classical light with measurements based on correlations similar to ghost imaging and the Hanbury-Brown and Twiss experiment \cite{h-t,texts}. This method has the advantage of being more robust with respect to losses \cite{thermal,thermal2}. Another approach is to build an interferometric lithography with use of classical coherent state \cite{coherent,coherent2}, which has similar setup to the case using entangled photon-number states.

In this paper we will consider improving spatial resolution beyond the Rayleigh diffraction limit using quantum imaging with an entangled photon-number state $|1,N\rangle$. In our imaging scheme by sending the $N$ degenerate photons to the object while keeping the non-degenerate photon and imaging lens in the laboratory, a factor of $N$ improvement can be achieved in spatial resolution enhancement compared to classical optics. The assumptions required for the enhancement by a factor of $N$ are that the $N$ photons sent to the object scatter off the same point and are detected by either an $N$-photon number detector or a bucket detector. This sub-Rayleigh imaging resolution may have important applications in such as improving sensitivities of classical sensors and remote sensing. We emphasize that it is the quantum nature of the state that offers such sub-wavelength resolving power with high visibility. However, the system is very sensitive to loss. While we give general results, our main concern will be with the case in which the object is far from the source and the detectors and optics are close to the source. A different but related approach to the one discussed here is given in \cite{giovannetti}.

We organize the paper as follows. We will discuss our imaging scheme with entangled photon-number state $|1,2\rangle$ in some detail in Sec.~II. In previous work \cite{wen1,wen2} we have shown that imaging occurs in correlation measurement, as in the ghost imaging case. Here we will show that under certain stringent conditions, the resolution
can be improved by a factor of $2$ compared to classical optics. In Sec.~III we generalize the scheme to the
$|1,N\rangle$ case and show that resolution improvement by a factor of $N$ can be obtained. In Sec.~IV some discussions will be addressed on other experimental configurations. Finally we will draw our conclusions in Sec.~V.  In an appendix we discuss the meaning of the approximation that the $N$ photons illuminate the same point on the object.

\section{Three-Photon Optics}
We start with three photons because this is the easiest case to investigate the various configurations. Throughout the paper we shall assume that the source of the three photons is a pure state and that the three-photon counting rate for three point detectors is give by
\begin{equation}
R_{cc}=\frac{1}{T^2}\int_{0}^{T}dt_1\int_{0}^{T}dt_2\int_{0}^{T}dt_3|\Psi(1,2,3)|^{2},\label{eq:Coin}
\end{equation}
where the three-photon amplitude is determined by matrix element between the vacuum state and the three-photon state
$|\psi\rangle$
\begin{equation}
\Psi(1,2,3)=\langle0|E^{(+)}_1E^{(+)}_2E^{(+)}_3|\psi\rangle,\label{eq:Ampl}
\end{equation}
and
\begin{equation}
E^{(+)}_j(\vec{\rho}_j,z_j,t_j)=\int{d}\omega_j\int{d^{2}}\alpha_jE_jf_j(\omega_j)e^{-i\omega_jt_j}g_j(\vec{\alpha}_j,
\omega_j;\vec{\rho}_j,z_j)a(\vec{\alpha}_j,\omega_j),\label{eq:freefield}
\end{equation}
where $E_j=\sqrt{\hbar\omega_{j}/2 \epsilon_{0}}$, $\vec{\alpha}_j$ is the transverse wave vector, and
$a(\vec{\alpha}_j,\omega_j)$ is a photon annihilation operator at the output surface of the source,
\begin{equation}
[a(\vec{\alpha},\omega),a^{\dagger}(\vec{\alpha}\prime,\omega\prime)]=\delta(\vec{\alpha}-\vec{\alpha}\prime)
\delta(\omega-\omega\prime).\label{eq:commutation rel}
\end{equation}
The function $f_{j}(\omega)$ is a narrow bandwidth filter function which is assumed to be peaked at $\Omega_j$. The function $g_j$ is the Green's function \cite{goodman,rubin} that describes the propagation of each mode from the output surface of the source to the $j$th detector at the transverse coordinate $\vec{\rho}_j$, at the distance from the
output surface of the crystal to the plane of the detector, $z_j$. $\Psi$ is referred to as the \textit{three-photon amplitude} (or three-photon wavefunction).

\begin{figure}[tbp]
\includegraphics[scale=0.6]{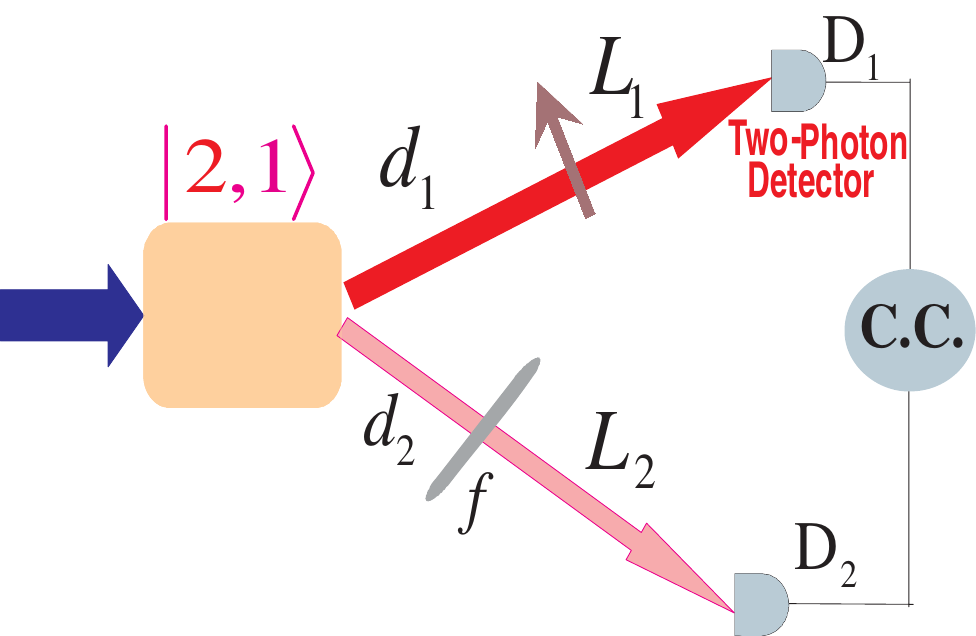}
\caption{(color online) Schematic of quantum imaging with a three-photon entangled state $|1,2\rangle$. $d_1$ is the distance from the output surface of the source to the object. $L_1$ is the distance from the object to a 2-photon detector, D$_1$. $d_2$ is the distance from the output surface of the source to the imaging lens with focal length $f$ and $L_2$ is the length from the imaging lens to a single-photon detector D$_2$, which scans coming signal photons in its transverse plane. ``C.C." represents the joint-detection measurement.}\label{fig:2photon}
\end{figure}

We start with the case in which the source produces three-photon entangled states with a pair of degenerate photons, that is $\psi\rightarrow\psi_{{1,2}}$
\begin{equation}
|\psi_{1,2}\rangle=\int{d}\omega_1{d}\omega_2\int{d^{2 }}\alpha_1d^2 \alpha_2\delta(2\omega_1+\omega_2-\Omega)\delta(
2\vec{\alpha}_1+\vec{\alpha}_2)a^{\dagger}(\vec{\alpha}_2,\omega_{2})\big[a^{\dagger}(\vec{\alpha}_1,\omega_{1})\big]^2
|0\rangle,\label{eq:state1}
\end{equation}
where $\Omega$ is a constant, $\omega_{1,2}$ and $\vec{\alpha}_{1,2}$ are the frequencies and transverse wave vectors of the degenerate and non-degenerate photons, respectively. The $\delta$-functions indicate that the source is assumed to produce three-photon states with perfect phase matching. We assume the paraxial approximation holds and that the temporal and transverse behavior of the waves factor. The frequency correlation determines the three-photon temporal
properties. The transverse momentum correlation determines the spatial properties of entangled photons. It is this wave-vector correlation that we are going to concentrate on. As discussed in \cite{wen1}, several imaging schemes can be implemented with this three-photon source. To demonstrate spatial resolution enhancement beyond the Rayleigh diffraction limit, consider the experimental setup shown in Fig.~\ref{fig:2photon}. It will be shown that for this configuration the spatial resolving power is improved by a factor of 2, provided the degenerate photons
illuminate the same point on the object and are detected by a two photon detector.
%%%%%%%%%%%%%%%%%%%%%%%%%%%%%%%%%%%%%%%%%%%%%%%%%%%
%%%%%%%%%%%%%%%%%%%%%%%%%%%%%%%%%%%%%%%%%%%%%%%%%%%

As depicted in Fig.~\ref{fig:2photon}, two degenerate photons with wavelength $\lambda_1$ are sent to a two-photon detector (D$_1$) after illuminating an object, and the non-degenerate photon with wavelength $\lambda_2$ propagates to a single-photon detector (D$_2$) after an imaging lens with focal length $f$. The three-photon amplitude
(\ref{eq:Ampl}) for detectors D$_1$ and D$_2$, located at $(z_1,\vec{\rho}_1)$ and $(z_2,\vec{\rho}_2)$, now is
\begin{equation}
\Psi\rightarrow\Psi_{1,2}=\langle0|E^{(+)}_2(\vec{\rho}_2,z_2,t_2)\big[E^{(+)}_1(\vec{\rho}_1,z_1,t_1)\big]^2|\psi_{1,2}
\rangle, \label{eq:coincidence}
\end{equation}
Following the treatments in \cite{rubin,goodman,wen1}, we evaluate the Green's functions
$g_1(\vec{\alpha}_1,\omega_2;\vec{\rho}_1,z_1)$ and $g_2(\vec{\alpha}_2,\omega_2;\vec{\rho}_2,z_2)$ for the experimental setup of Fig.~\ref{fig:2photon} assuming that the narrow bandwidth filter allows us to make the assumption that $\omega_{j}=\Omega_j+\nu_j$ where $|\nu_j|\ll\Omega_j$ and $2\Omega_1+\Omega_2=\Omega$.

In the paraxial approximation it is convenient to write
\begin{equation}
g_j(\vec{\alpha}_j,\omega_j;\vec{\rho}_j,z_j)=\frac{\omega_je^{i\omega_jz_j /c}}{i 2\pi cL_jd_j}
\chi_j(\vec{\alpha}_{j},\omega_{j};\vec{\rho}_{j},z_{j}),\label{eq:chi}
\end{equation}
then
\begin{eqnarray}
\chi_1(\vec{\alpha}_1,\Omega_1;\vec{\rho}_1,z_1)&=&e^{-i\frac{d_1|\vec{\alpha}_1|^2}{2K _1}}\int{d^{2}}\rho_o
A(\vec{\rho}_o)e^{i\frac{K_1|\vec{\rho}_o|^2}{2L_1}}e^{-i\frac{K_1\vec{\rho}_1\cdot\vec{\rho}_o}{L_1}}
e^{i\vec{\alpha}_1\cdot\vec{\rho}_o},\label{eq:g1}\\
\chi_2(\vec{\alpha}_2,\Omega_2;\vec{\rho}_2,z_2)&=&e^{-i\frac{d_2|\vec{\alpha}_2|^2}{2K_2}}\int{d^{2}}\rho_l
e^{i\frac{K_2|\vec{\rho}_l|^2}{2}(\frac{1}{L_2}-\frac{1}{f})}e^{i(\vec{\alpha}_2-\frac{K_2}{L_2}\vec{\rho}_2)\cdot
\vec{\rho}_l},\label{eq:g2}
\end{eqnarray}
where we replace $\omega_j$ by $\Omega_j$ in $\chi_j$, $K_j=\Omega_j /c=2\pi/\lambda_j$, $z_1=d_1+L_1$, and $z_2=d_2+L_2$, respectively. In Eqs.~(\ref{eq:g1}) and (\ref{eq:g2}), $A(\vec{\rho}_o)$ is the aperture function of the object, and $\vec{\rho}_o$ and $\vec{\rho}_l$ are two-dimensional vectors defined, respectively, on the object and the imaging lens planes. With use of Eqs.~(\ref{eq:freefield}) and (\ref{eq:state1}), the three-photon amplitude (\ref{eq:coincidence}) becomes
\begin{equation}
\Psi_{1,2}=e^{i(2\Omega_1\tau_1+\Omega_2\tau_2)}\Phi_{1,2},\label{Phi12}
\end{equation}
where $\tau_j=t_j-z_j/c$ and
\begin{eqnarray}
\Phi_{1,2}=\int{d}\nu_1d\nu_2\delta(2\nu_1+\nu_2)e^{i(2\nu_1 \tau_1+\nu_2\tau_2)}f_1(\Omega_1+\nu_1)^2f_2 (\Omega_2+\nu_2)B_{1,2}.\label{eq:Psi1}
\end{eqnarray}
where
\begin{eqnarray}
B_{1,2}&=&B_{0}\int{d^{2}}\rho_oA(\vec{\rho}_o)e^{i\frac{K_1|\vec{\rho}_o|^2}{2L_1}}e^{-i\frac{K_1\vec{
\rho}_1\cdot\vec{\rho}_o}{L_1}}\int{d^{2}}\rho'_oA(\vec{\rho}'_o)e^{i\frac{K_1|\vec{\rho}'_o|^2}{2L_1}}e^{-i
\frac{K_1\vec{\rho}_1\cdot\vec{\rho}'_o}{L_1}}\int{d^{2}}\rho_le^{i\frac{K_2|\vec{\rho}_l|^2}{2}(\frac{1}{
L_2}-\frac{1}{f})}e^{-i\frac{K_2}{L_2}\vec{\rho}_2\cdot\vec{\rho}_l}\nonumber\\
&&\times\int{d^{2 }}\alpha_1e^{-i|\vec{\alpha}_1|^2(\frac{d_1}{K_1}+\frac{2d_2}{K_2})}e^{-i\vec{\alpha}_1
\cdot(2\vec{\rho}_l-\vec{\rho}_o-\vec{\rho}'_o)},\label{eq:psiapproximation}
\end{eqnarray}
where we collect all the slowly varying quantities into the constant $B_{0}$. To proceed the discussion, in the following we will consider two different detection schemes. One uses a point two-photon detector for two degenerate photons after the object and the other has a two-photon bucket detector.

\subsection{Point Two-Photon Detector Scheme}
In this detection scheme, a point two-photon detector is necessary to retrieve the information of degenerate photons scattered off the same point in the object. We therefore make the key assumption that the detector D$_1$ is only sensitive to the signals from the same point in the object, i.e., $\delta(\vec{\rho}_o-\vec{\rho}'_o)$ [The validity of this assumption is addressed in the Appendix]. With this assumption, Eq.~(\ref{eq:psiapproximation}) becomes
\begin{eqnarray}
B_{1,2}&=&B_{0}\int{d^{2}}\rho_oA^2(\vec{\rho}_o)e^{i\frac{K_1|\vec{\rho}_o|^2}{L_1}}e^{-i\frac{2K_1\vec{
\rho}_1\cdot\vec{\rho}_o}{L_1}}\int{d^{2}}\rho_le^{i\frac{K_2|\vec{\rho}_l|^2}{2}(\frac{1}{L_2}-\frac{1}{f})}
e^{-i\frac{K_2}{L_2}\vec{\rho}_2\cdot\vec{\rho}_l}\nonumber\\
&&\times \int{d^{2 }}\alpha_1e^{-i|\vec{\alpha}_1|^2(\frac{d_1}{K_1}
+\frac{2d_1}{K_2})}e^{-2i\vec{\alpha}_1\cdot(\vec{\rho}_l-\vec{\rho}_o)}.\label{eq:Psiassumption}
\end{eqnarray}
Completing the integration on the transverse mode $\vec{\alpha}_1$ in Eq.~(\ref{eq:Psiassumption}) gives
\begin{eqnarray}
B_{1,2}&=& B_{0}\int{d^{2}}\rho_oA^2(\vec{\rho}_o)e^{iK_1|\vec{\rho}_o|^2
[\frac{1}{L_1}+\frac{1}{d_1+(2\lambda_2/\lambda_1)d_2}]}e^{-i\frac{2K_1\vec{\rho}_1\cdot\vec{\rho}_o}{L_1}}\nonumber\\
&&\times\int{d^{2}}\rho_le^{i\frac{K_2|\vec{\rho}_l|^2}{2}[\frac{1}{L_2}+\frac{1}{d_2+(\lambda_1/2\lambda_2)d_1
}-\frac{1}{f}]}e^{-iK_2\vec{\rho}_l\cdot[\frac{\vec{\rho}_2}{L_2}+\frac{\vec{\rho}_o}{d_2+(\lambda_1/2
\lambda_2)d_1}]}.\label{eq:psiassumption2}
\end{eqnarray}
By imposing the Gaussian thin-lens imaging condition in Eq.~(\ref{eq:psiassumption2})
\begin{eqnarray}
\frac{1}{f}=\frac{1}{L_2}+\frac{1}{d_2+(\lambda_1/2\lambda_2)d_1},\label{eq:GTLE1}
\end{eqnarray}
the transverse part of the three-photon amplitude reduces to
\begin{eqnarray}
B_{1,2}&=&B_{0}\int{d^{2}}\rho_oA^2(\vec{\rho}_o)e^{iK_1|\vec{\rho}_o|^2[\frac{1}{L_1}+\frac{1}{d_1
+(2\lambda_2/\lambda_1)d_2}]}e^{-i\frac{2K_1\vec{\rho}_1\cdot\vec{\rho}_o}{L_1}}\mathbf{somb}\bigg(\frac{2\pi{R}}{
\lambda_2[d_2+(\lambda_1/2\lambda_2)d_1]}\bigg|\vec{\rho}_o+\frac{\vec{\rho}_2}{m}\bigg|\bigg),\label{eq:psif}
\end{eqnarray}
where $R$ is the radius of the imaging lens, $R/[d_2+(\lambda_1/2\lambda_2)d_1]$ may be thought of as the numerical aperture of the imaging system, and $m=L_2/[d_2+(\lambda_1/2\lambda_2)d_1]$ is the magnification factor. In Eq.~(\ref{eq:psif}) the Airy disk is determined, as usual, by $\mathbf{somb}(x)=2J_1(x)/x$, where $J_1(x)$ is the first-order Bessel function.

Before proceeding with the discussion of resolution, let us look at the physics behind Eqs.~(\ref{eq:GTLE1}) and (\ref{eq:psif}). Equation (\ref{eq:GTLE1}) defines the image plane where the ideal the point-to-point mapping of the
object plane occurs. The unique point-to-point correlation between the object and the imaging planes is the result
of the transverse wavenumber correlation and the fact that we have assumed that the degenerate photons illuminate the same object point. Let us make a comparison with the two-photon and three-photon geometrical optics
\cite{rubin,wen1,rubin2008}. In the Gauss thin lens equation the distance between the imaging lens and the object planes, $d_2+(\lambda_1/2\lambda_2)d_1$ is similar to the form that appears in the non-degenerate two-photon case except for the factor of 2. This factor 2 comes from the degeneracy of the pair of photons that illuminate the object. As we will show below, this factor of 2 is the source of the improved spatial resolution. Equation (\ref{eq:psif}) implies that a coherent and inverted image magnified by a factor of m is produced in the plane of $D_{2}$. Of course, there really is no such image and the true image is \textit{nonlocal}. The point-spread function in Eq.~(\ref{eq:psif}) is generally determined by both wavelengths of the degenerate and non-degenerate photons.

To examine the resolution using the Rayleigh criterion, we consider an object consisting of two point scatters, one located at the origin and the other at the point $\vec{a}$ in the object plane,
\begin{equation}
A(\vec{\rho}_o)^2=A_0^2\delta(\vec{\rho}_o)+A_{\vec{a}}^2\delta(\vec{\rho}_o-\vec{a}).\label{eq:object}
\end{equation}
By substituting Eq.~(\ref{eq:object}) into (\ref{eq:psif}) we obtain
\begin{eqnarray}
B_{1,2}=B_{0}\bigg({A}^2_0 \mathbf{somb}\bigg(\frac{2\pi{R}}{\lambda_2}\bigg|\frac{\vec{\rho}_2}{L_2}\bigg|\bigg)+e^{i
\varphi_2}A_{\vec{a}}^2\mathbf{somb}\bigg[\frac{2\pi{R}}{\lambda_2}\bigg|\frac{\vec{\rho}_2}{L_2}+\frac{\vec{a}}{d_2+(
\lambda_1/2\lambda_2)d_1}\bigg|\bigg]\bigg),\label{eq:Psirayleigh}
\end{eqnarray}
where the phase
\begin{equation}
\varphi_2=K_1\bigg[|\vec{a}|^2\bigg(\frac{1}{L_1}+\frac{1}{d_1+d_2(2\lambda_2/\lambda_1)}\bigg)-\frac{\vec{a}
\cdot(\vec{\rho}_1+\vec{\rho}'_1)}{L_1}\bigg]\label{eq:phase2}
\end{equation}
indicates that the image is coherent. For a point $2$-photon detector, we require $\vec{\rho}_1=\vec{\rho}'_1$ in Eq.~(\ref{eq:phase2}). As is well-known \cite{goodman} for coherent imaging the Rayleigh criterion is not the best choice for characterizing the resolution, however, it is indicative of the resolution that can be attained and it is convenient. For a circular aperture, the radius of the Airy disk, $\xi$, is determined by the point-spread function, which is
\begin{equation}
\xi=0.61\frac{\lambda_2L_2}{R}.\label{eq:airydisk1}
\end{equation}
Note that the radius of the Airy disk is proportional to the wavelength of the non-degenerate photon. This is the standard result as obtained in classical optics. Using the Rayleigh criterion, the image of the second term in Eq.~(\ref{eq:Psirayleigh}) is taken to lie on the edge of the Airy disk of the first term, therefore,
\begin{equation}
a_{\mathrm{m}}=0.61\frac{\lambda_2}{R}\bigg(d_2+\frac{\lambda_1}{2\lambda_2}d_1\bigg).\label{eq:resolution1}
\end{equation}
We see from Eq.~(\ref{eq:resolution1}) that the resolution depends on the wavelengths of the degenerate and the non-degenerate photons. In the case that $d_1\gg{d}_2$, so that $d_2+(\lambda_1/2\lambda_2)d_1$, is approximately  $(\lambda_1/2\lambda_2)d_1$. In this case Eq.~(\ref{eq:GTLE1}) implies that $L_2\approx f$ and the radius of the Airy disk approaches to $1.22\lambda_2f/R$, and
\begin{equation}
a_{\mathrm{m}}=0.61\frac{\lambda_1d_1/2}{R}.\label{eq:resolution2}
\end{equation}
Equation~(\ref{eq:resolution2}) shows a gain in spatial resolution of a factor of 2 compared to classical optics.
Furthermore, there is no background term which is characteristic of the quantum case.

\subsection{Bucket Detector Scheme}
If the two-photon detector is replaced by a bucket detector and  the two degenerate photons are collected by two single-photon detection events, located at $(L_1,\vec{\rho}_1)$ and $(L_1,\vec{\rho}'_1)$, in the bucket, Eq.~(\ref{eq:psiapproximation}) becomes
\begin{eqnarray}
B_{1,2}&=&B_{0}\int{d^{2}}\rho_oA(\vec{\rho}_o)e^{i\frac{K_1|\vec{\rho}_o|^2}{2L_1}}e^{-i\frac{K_1\vec{
\rho}_1\cdot\vec{\rho}_o}{L_1}}\int{d^{2}}\rho'_oA(\vec{\rho}'_o)e^{i\frac{K_1|\vec{\rho}'_o|^2}{2L_1}}e^{-i
\frac{K_1\vec{\rho}'_1\cdot\vec{\rho}'_o}{L_1}}\int{d^{2}}\rho_le^{i\frac{K_2|\vec{\rho}_l|^2}{2}(\frac{1}{
L_2}-\frac{1}{f})}e^{i\frac{K_2}{L_2}\vec{\rho}_2\cdot\vec{\rho}_l}\nonumber\\
&&\times\int{d^{2 }}\alpha_1e^{-i|\vec{\alpha}_1|^2(\frac{d_1}{K_1}+\frac{2d_2}{K_2})}e^{-i\vec{\alpha}_1
\cdot(2\vec{\rho}_l-\vec{\rho}_o-\vec{\rho}'_o)}.\label{eq:bucketdetector}
\end{eqnarray}
Under the assumption that the  two degenerate photons are scattered off the same point in the object, Eq.~(\ref{eq:bucketdetector}) takes the similar form as Eq.~(\ref{eq:Psiassumption}), except that the second phase term in the first integrand of (\ref{eq:Psiassumption}) is replaced by $\mathrm{exp}\big[-i\frac{K_1(\vec{\rho}_1+\vec{\rho}'_1)\cdot\vec{\rho}_o}{L_1}\big]$.
It is easy to show that the Gaussian thin-lens equation takes the same form as Eq.~(\ref{eq:GTLE1}). By performing the same analysis as done in Sec.~IIA on the resolving two spatially close point scatters, the three-photon amplitude (\ref{eq:Psirayleigh}) now is
\begin{eqnarray}
B_{1,2}=B_{0}\bigg(A_0^2\mathbf{somb}\bigg(\frac{2\pi{R}}{\lambda_2}\bigg|\frac{\vec{\rho}_2}{L_2}\bigg|\bigg)+e^{i
\varphi_2}A_{\vec{a}}^2\mathbf{somb}\bigg[\frac{2\pi{R}}{\lambda_2}\bigg|\frac{\vec{\rho}_2}{L_2}+\frac{\vec{a}}{d_2+(
\lambda_1/2\lambda_2)d_1}\bigg|\bigg]\bigg).\label{eq:Psirayleighbucket}
\end{eqnarray}
Since the bucket detector gives no position information, we must square the amplitude and integrating over the bucket detector,
\begin{equation}
I=\int{d^{2}}\rho_{1}  \int{d^{2}}\rho_{1}'  |B_{1,2}|^{2}=s_{b}^2|B_{0}|^{2}\bigg(|A_0|^4 \mathbf{somb}^{2}\bigg(\frac{2\pi{R}}{\lambda_2}\bigg|\frac{\vec{\rho}_2}{L_2}\bigg|\bigg)+
|A_{\vec{a}}|^{4} \mathbf{somb}^2\bigg[\frac{2\pi{R}}{\lambda_2}\bigg|\frac{\vec{\rho}_2}{L_2}+\frac{\vec{a}}{d_2+(
\lambda_1/2\lambda_2)d_1}\bigg|\bigg]\bigg)\label{eq:bucket}
\end{equation}
where $s_b$ is the area of the bucket detector.  It is easy to see that the spatial resolution improvement is the same as in Sec.~IIA, the difference is that now we get an incoherent image.  The advantage is that a two photon bucket detector should be easier to construct than a point two photon detector.

\section{$N+1$ Photon Optics}
%%%%%%%%%%%%%%%%%%%%%%%%%%%%%%%%%%%%%%%%%%%%%%%%%%%
\begin{figure}[tbp]
\includegraphics[scale=0.6]{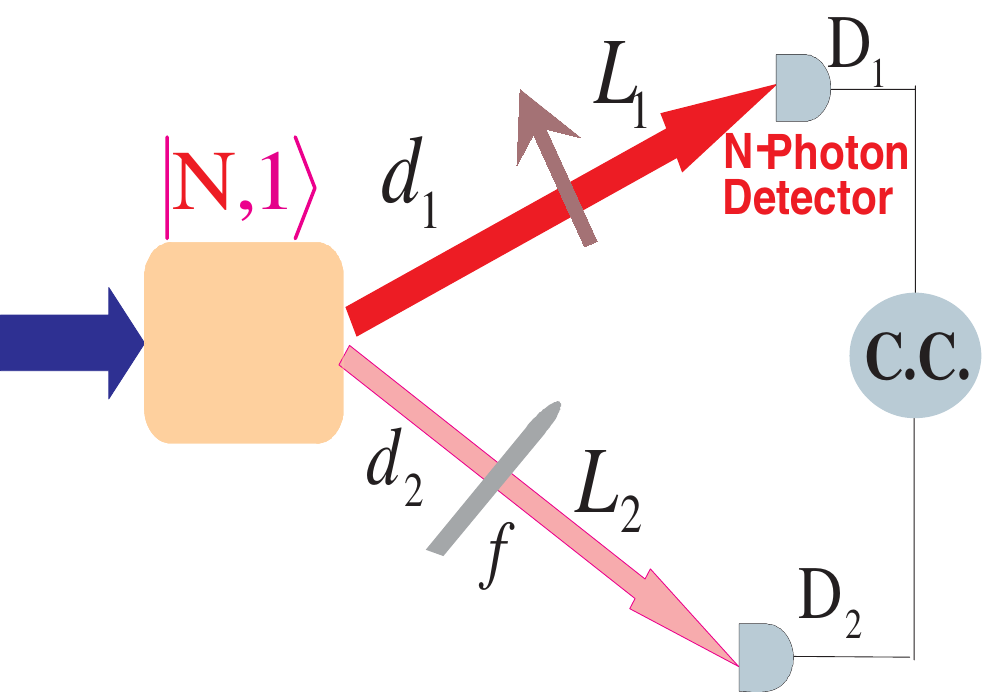}
\caption{(color online) Generalization of quantum imaging with $N+1$ entangled photons in state $|1,N\rangle$. For notations please refer to Fig.~\ref{fig:2photon} except that here D$_1$ is an $N$-photon detector. The image is formed in the coincidence measurement and is not localized at either detector.}\label{fig:Nphoton}
\end{figure}
%%%%%%%%%%%%%%%%%%%%%%%%%%%%%%%%%%%%%%%%%%%%%%%%%%%
In Sec.~II, we have shown that with the entangled photon-number state $|1,2\rangle$, the ability to resolve two point sources in the object can be improved by a factor of 2 by sending two degenerate photons to the object while keeping the non-degenerate photon and imaging lens in the laboratory. In this section, we are going to generalize the experimental configuration (Fig.~\ref{fig:2photon}) with use of the entangled state of $|1,N\rangle$, as described in Fig.~\ref{fig:Nphoton}. For simplicity, we first address the case shown in Fig.~\ref{fig:Nphoton} where the $N$ degenerate photons traverse to the $N$-photon detector, D$_1$, after the object and the non-degenerate photon propagates to the single-photon detector, D$_2$. The assumption required for the enhancement by a factor of $N$ are that the $N$ photons sent to the object scatter off the same point and are detected by the $N$-photon detector, D$_1$.

The $N+1$ photons are assumed to be in a non-normalized pure state
\begin{eqnarray}
|\psi_{1,N}\rangle=\int{d}\omega_1{d}\omega_2\int{d^{2 }}\alpha_1d^2\alpha_2\delta(N\omega_1+\omega_2-\Omega)\delta
(N\vec{\alpha}_1+\vec{\alpha}_2)a^{\dagger}_{\vec{k}_2}\big(a^{\dagger}_{\vec{k}_1}\big)^N|0\rangle.\label{eq:state2}
\end{eqnarray}
Again the $\delta$-functions in Eq.~(\ref{eq:state2}) indicate perfect phase matching. The $N+1$-photon coincidence counting rate is defined as
\begin{eqnarray}
R_{cc}=\frac{1}{T}\int^T_0dt_1\int^T_0dt_2\cdots\int^T_0dt_{N+1}|\Psi_{1,N}(1,2,\cdots,N+1)|^2,\label{eq:coincidence2}
\end{eqnarray}
where $\Psi_{1,N}$ is referred to as the \textit{$N+1$-photon amplitude}. That is
\begin{eqnarray}
\Psi_{1,N}(1,2,\cdots,N+1)&=&\langle0|E^{(+)}_1E^{(+)}_2\cdots{E}^{(+)}_{N+1}|\psi_{1,N}\rangle\nonumber\\
&=&\langle0|E^{(+)}_2(\vec{\rho}_2,z_2,t_2)[E^{(+)}_1(\vec{\rho}_1,z_1,t_1)]^N|\psi_{1,N}\rangle.\label{eq:ampN}
\end{eqnarray}

Following the procedure done for the $|1,2\rangle$ case, we calculate the transverse part of the $N+1$-photon amplitude $\Psi_{1,N}$ (\ref{eq:ampN}) as
\begin{eqnarray}
\Psi_{1,N}&=&e^{i(N\Omega_1\tau_1+\Omega_2\tau_2)}\Phi_{1,N}(\tau_1,\tau_2)B_{1,N}\nonumber\\
B_{1,N}&=&B_0\underbrace{\int{d^{2}}\rho_oA(\vec{\rho}_o)e^{i\frac{K_1|\vec{\rho}_o|^2}{2L_1}}e^{-i\frac{
K_1\vec{\rho}_1\cdot\vec{\rho}_o}{L_1}}\cdots\int{d^{2}}\rho'_oA(\vec{\rho}'_o)e^{i\frac{K_1|\vec{\rho}'_o|
^2}{2L_1}}e^{-i\frac{K_1\vec{\rho}_1\cdot\vec{\rho}'_o}{L_1}}}_{\mathrm{N\;fold}}\int{d^{2}}\rho_le^{i\frac{
K_2|\vec{\rho}_l|^2}{2}(\frac{1}{L_2}-\frac{1}{f})}e^{-i\frac{K_2}{L_2}\vec{\rho}_2\cdot\vec{\rho}_l}\nonumber\\
&&\times\int{d^{2}}\alpha_1e^{-i\frac{N^2|\vec{\alpha}_1|^2}{2}(\frac{d_1}{NK_1}+\frac{d_2}{K_2})}e^{-i\vec{\alpha}_1
\cdot(N\vec{\rho}_l-\underbrace{\vec{\rho}_o-\cdots-\vec{\rho}'_o}_{\mathrm{N}})}.\label{eq:psiapproximation2}
\end{eqnarray}
Here $\Phi_{1,N}(\tau_1,\tau_2)$ describes the temporal behavior of entangled three photons. By applying the same argument that the $N$-photon detector D$_1$ only receives the signals from the same spatial point in the object, Eq.~(\ref{eq:psiapproximation2}) can be further simplified as
\begin{eqnarray}
B_{1,N}&=&B_0\int{d^{2}}\rho_oA^N(\vec{\rho}_o)e^{i\frac{NK_1|\vec{\rho}_o|^2}{2L_1}}e^{-i\frac{N K_1
\vec{\rho}_1\cdot\vec{\rho}_o}{L_1}}\int{d^{2}}\rho_le^{i\frac{K_2|\vec{\rho}_l|^2}{2}(\frac{1}{L_2}-\frac{1}{
f})}e^{-i\frac{K_2}{L_2}\vec{\rho}_2\cdot\vec{\rho}_l}\nonumber\\
&&\times \int{d^{2}}\alpha_1e^{-i\frac{N^2|\vec{\alpha}_1|^2}{2}(\frac{d_1}{N
K_1}+\frac{d_2}{K_2})}e^{-Ni\vec{\alpha}_1\cdot(\vec{\rho}_l-\vec{\rho}_o)}.\label{eq:Psiassumption2}
\end{eqnarray}
Performing the integration on the transverse mode $\vec{\alpha}_1$ in Eq.~(\ref{eq:Psiassumption2}) gives
\begin{eqnarray}
B_{1,N}&=&B_0\int{d^{2}}\rho_oA^N(\vec{\rho}_o)e^{i\frac{NK_1|\vec{\rho}_o|^2}{2}[\frac{1}{L_1}+\frac{1}{d_1+(N
\lambda_2/\lambda_1)d_2}]}e^{-i\frac{NK_1\vec{\rho}_1\cdot\vec{\rho}_o}{L_1}}\nonumber\\
&&\times\int{d^{2}}\rho_le^{i\frac{K_2|\vec{\rho}_l|^2}{2}[\frac{1}{L_2}+\frac{1}{d_2+(\lambda_1/N\lambda_2)d_1
}-\frac{1}{f}]}e^{-iK_2\vec{\rho}_l\cdot[\frac{\vec{\rho}_2}{L_2}+\frac{\vec{\rho}_o}{d_2+(\lambda_1/N\lambda_2)d_1}
]},\label{eq:psiassumptionn}
\end{eqnarray}
where, again, we have assumed multimode generation in the process. Applying the Gaussian thin-lens imaging condition
\begin{eqnarray}
\frac{1}{f}=\frac{1}{L_2}+\frac{1}{d_2+(\lambda_1/N\lambda_2)d_1},\label{eq:GTLEn}
\end{eqnarray}
the transverse part of the $N+1$-photon amplitude (\ref{eq:psiassumptionn}) between detectors D$_1$ and D$_2$ now
becomes
\begin{eqnarray}
B_{1,N}&=&B_0\int{d^{2}}\rho_oA^N(\vec{\rho}_o)e^{i\frac{NK_1|\vec{\rho}_o|^2}{2}[\frac{1}{L_1}+\frac{1}{d_1+(N\lambda
_2/\lambda_1)d_2}]}e^{-i\frac{NK_1\vec{\rho}_1\cdot\vec{\rho}_o}{L_1}}\mathbf{somb}\bigg[\frac{2\pi{R}}{\lambda_2}\bigg|
\frac{\vec{\rho}_2}{L_2}+\frac{\vec{\rho}_o}{d_2+(\lambda_1/N\lambda_2)d_1}\bigg|\bigg].
\label{eq:psin}
\end{eqnarray}
As expected, Eqs.~(\ref{eq:GTLEn}) and (\ref{eq:psin}) have the similar forms as Eqs.~(\ref{eq:GTLE1}) and (\ref{eq:psif}) for the $|1,2\rangle$ case. The unique point-to-point relationship between the object and the imaging planes is enforced by the Gaussian thin-lens equation (\ref{eq:GTLEn}). The coherent and inverted image is demagnified by a factor of $L_2/[d_2+d_1(\lambda_1/N\lambda_2)]$. The spatial resolution is determined by the width of the point-spread function in Eq.~(\ref{eq:psin}). Note that a factor of $N$ appears in the distance between the imaging lens and the object planes, $d_2+d_1(\lambda_1/N\lambda_2)$. We emphasize again that the image is nonlocal and exists in the coincidence events.

To study the spatial resolution, we again consider the object represented by Eq.~(\ref{eq:object}). Plugging Eq.~(\ref{eq:object}) into (\ref{eq:psin}) yields
\begin{eqnarray}
B_{1,N}=B_0\bigg(A_{0}^N\mathbf{somb}\bigg(\frac{2\pi{R}}{\lambda_2}\bigg|\frac{\vec{\rho}_2}{L_2}\bigg|\bigg)+e^{i
\varphi_{N}}A_{\vec{a}}^N\mathbf{somb}\bigg[\frac{2\pi{R}}{\lambda_2}\bigg|\frac{\vec{\rho}_2}{L_2}+\frac{\vec{a}}{d_2+
(\lambda_1/N\lambda_2)d_1}\bigg|\bigg]\bigg).\label{eq:Psinrayleigh}
\end{eqnarray}
For $N$ single photon detectors located at $\vec{\rho}_{1}^{(1)},\cdots,\vec{\rho}_1^{(N)}$ the phase is given by
\begin{equation}
\varphi_N=K_1\bigg[\frac{N|\vec{a}|^2}{2}\bigg(\frac{1}{L_1}+\frac{1}{d_1+d_2(N\lambda_2/\lambda_1)}\bigg)-\frac{
\vec{a}\cdot(\overbrace{\vec{\rho}_{1}^{(1)}+\vec{\rho}^{(2)}_{1}+\cdots}^{\mathrm{N}})}{L_1}\bigg]\label{eq:phaseN}
\end{equation}
For a point $N$-photon number detector, we require $\vec{\rho}_{1}^{(1)}=\vec{\rho}^{(2)}_{1}=\cdots$ and a coherent imaging is achievable in this case. The first term on the right-hand side in Eq.~(\ref{eq:Psirayleigh}) gives the radius of the Airy disk, which is the same as the $|1,2\rangle$ case, see Eq.~(\ref{eq:airydisk1}). Applying the Rayleigh criterion, the minimum resolvable distance between two points in the transverse plane now is
\begin{eqnarray}
a_{\mathrm{m}}=0.61\frac{\lambda_2}{R}\bigg(d_2+\frac{\lambda_1}{N\lambda_2}d_1\bigg).\label{eq:resolutionn}
\end{eqnarray}
For the case of $N=2$, Eq.~(\ref{eq:resolutionn}) reduces to Eq.~(\ref{eq:resolution1}). In the case that $d_1\gg{d_2}$, this becomes
\begin{eqnarray}
a_{\mathrm{m}}=0.61\frac{\lambda_1d_1}{NR}.\label{eq:resolutionf}
\end{eqnarray}
As expected, Eq.~(\ref{eq:resolutionf}) shows a gain in sub-Rayleigh resolution by a factor of $N$ with respect to what one would obtain in classical optics. We therefore conclude that in the proposed imaging protocol, the spatial resolving power can be improved by a factor of $N$ with use of the entangled photon-number state $|1,N\rangle$. Furthermore, because we are using an entangled state with a specific type of detector, the image has high contrast because of the lack of background noise.

By following the analysis in Sec.~IIB, we can show that by replacing the $N$-photon detector with an $N$-photon bucket detector, we get an incoherent image but the sub-Rayleigh imaging process is not changed.

\section{Discussions and other Configurations}
In the previous two sections, we have analyzed a novel ghost imaging by sending $N$ degenerate photons to the object while keeping the non-degenerate photon and imaging lens in the lab. We find that if the distance between the object plane and the output surface of the source is much greater than the distance between the imaging lens and the single-photon detector planes, we can gain spatial resolution improvement in the object by a factor of $N$ compared to classical optics. In the cases that we have discussed in this paper, this enhancement beyond the Rayleigh criterion is due to the quantum nature of the entangled photon-number state. The assumptions required for such an enhancement are that the $N$ degenerate photons sent to the object scatter off the same point and are detected by either an $N$-photon number detector or a bucket detector. An $N$-photon bucket detector is much easier to realize than an $N$-photon point detector. Such a bucket detector could be an array of single photon point detectors which only sent a signal to the coincidence circuit if exactly $N$ of them fired.

%%%%%%%%%%%%%%%%%%%%%%%%%%%%%%%%%%%%%%%%%%%%%%%%%%%
\begin{figure}[tbp]
\includegraphics[scale=0.6]{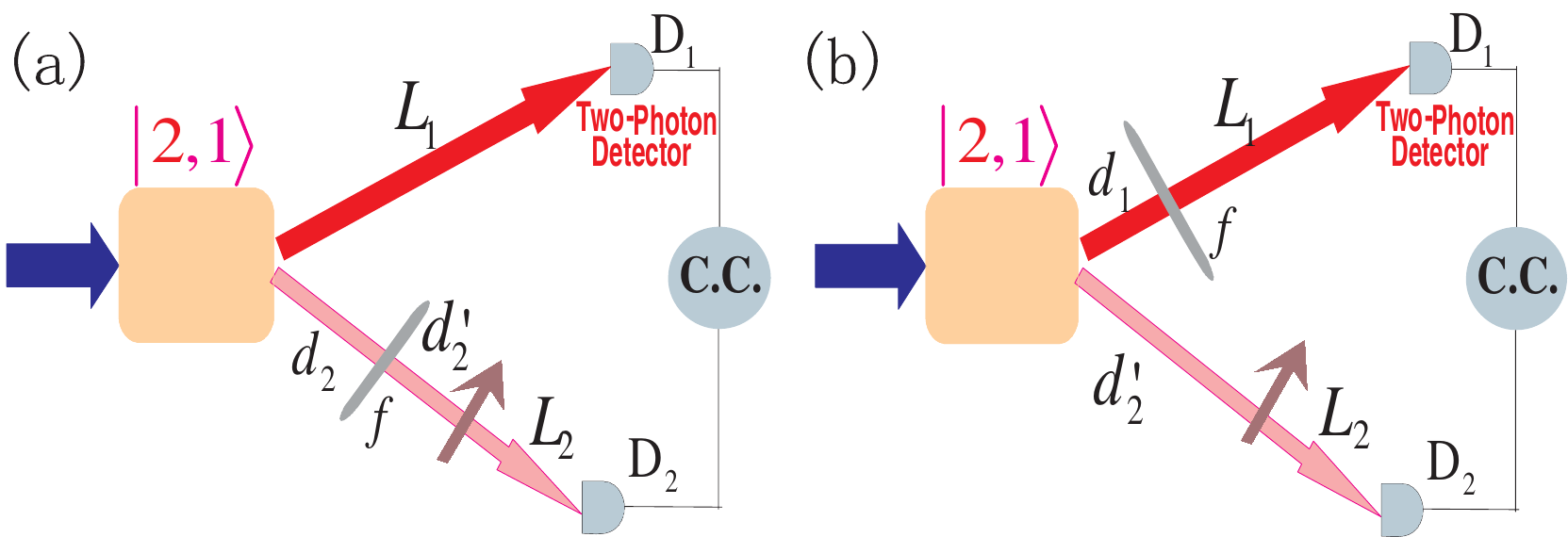}
\caption{(color online) Other schematics of quantum ghost imaging with three entangled photons in state $|1,2\rangle$. (a) Both the imaging lens and the object are inserted in the non-degenerate photon channel. (b) The imaging lens is placed in the degenerate photon pathway while the object is in the non-degenerate optical pathway.}
\label{fig:otherconfigurations}
\end{figure}
%%%%%%%%%%%%%%%%%%%%%%%%%%%%%%%%%%%%%%%%%%%%%%%%%%%
Besides the favorable configuration discussed above, one may wonder what happens if we switch the $N$ degenerate photons to detector $D_1$ and the non-degenerate photon to $D_2$ after an imaging lens and an object? Do we gain any spatial resolution improvement? To answer the questions, let us look at the $|1,2\rangle$ case as illustrated in Fig.~\ref{fig:otherconfigurations}(a). Following the treatments in Sec.~IIA, after some algebra we find that the transverse part of the three-photon amplitude (\ref{eq:coincidence}) is
\begin{eqnarray}
B_{1,2}&=&B_0 \int{d^{2}}\rho_oA(\vec{\rho}_o)e^{i\frac{K_2|\vec{\rho}_o|^2}{2}(\frac{1}{L_2}+\frac{1}{d'_2})}
e^{-i\frac{K_2\vec{\rho}_2\cdot\vec{\rho}_o}{L_2}}\nonumber\\
&&\times\int{d^{2}}\rho_le^{i\frac{K_2|\vec{\rho}_l|^2}{2}[\frac{1}{d'_2}+\frac{1}{d_2+(\lambda_1/2\lambda_2)L_1}-
\frac{1}{f}]}e^{-iK_2\vec{\rho}_l\cdot[\frac{\vec{\rho}
_o}{d'_2}+\frac{\vec{\rho}_1}{d_2+(\lambda_1/2\lambda_2)L_1}]}.\label{eq:Psi22}
\end{eqnarray}
In the derivation of Eq.~(\ref{eq:Psi22}), the Green's functions associated with each beam give
\begin{eqnarray}
\chi_1(\vec{\alpha}_1,\Omega_1;\vec{\rho}_1,L_1)&=&e^{-i\frac{L_1|\vec{\alpha}_1|^2}{2K_1}}e^{i\vec{\rho}_1\cdot
\vec{\alpha}_1},\label{eq:g12}\nonumber\\
\chi_2(\vec{\alpha}_2,\Omega_2;\vec{\rho}_2,z_2)&=&e^{-i\frac{d_2|\vec{\alpha}_2|^2}{2K_2}}\int{d^{2}}\rho_oA(
\vec{\rho}_o)e^{i\frac{K_2|\vec{\rho}_o|^2}{2}(\frac{1}{L_2}+\frac{1}{d'_2})}e^{-i\frac{K_2\vec{\rho}_2\cdot
\vec{\rho}_o}{L_2}}\int{d^{2}}\rho_le^{i\frac{K_2|\vec{\rho}_l|^2}{2}(\frac{1}{d'_2}-\frac{1}{f})}e^{i\vec{\rho}
_l\cdot(\vec{\alpha}_2-\frac{K_2\vec{\rho}_o}{d'_2})}.\nonumber
\end{eqnarray}
Applying the Gaussian thin-lens imaging condition
\begin{eqnarray}
\frac{1}{d'_2}+\frac{1}{d_2+(\lambda_1/2\lambda_2)L_1}=\frac{1}{f},\label{eq:GTLE2}
\end{eqnarray}
the transverse spatial part of the three-photon amplitude (\ref{eq:Psi22}) reduces to
\begin{eqnarray}
B_{1,2}=B_0\int{d^{2}}\rho_oA(\vec{\rho}_o)e^{i\frac{K_2|\vec{\rho}_o|^2}{2}(\frac{1}{L_2}+\frac{1}{d'_2})}
e^{-i\frac{K_2\vec{\rho}_2\cdot\vec{\rho}_o}{L_2}}\mathbf{somb}\bigg(\frac{2\pi{R}}{\lambda_2}\bigg|\frac{\vec{
\rho}_o}{d'_2}+\frac{\vec{\rho}_1}{d_2+(\lambda_1/2\lambda_2)L_1}\bigg|\bigg).\label{eq:Psi222}
\end{eqnarray}
From this we see that the magnification is $m=[d_2+(\lambda_1/2\lambda_2)L_1]/d'_2$.  Comparing Eqs.~(\ref{eq:GTLE2}) and (\ref{eq:Psi222}) with Eqs.~(\ref{eq:GTLE1}) and (\ref{eq:psif}), we see that the distances between the object and the thin lens and between the thin lens and the imaging plane are interchanged. Since the degenerate photons are measured at the imaging plane in the setup of Fig.~\ref{fig:otherconfigurations}(a), the requirement of a point $N$-photon detector cannot be relaxed.

Computing the spatial resolution as in Sec. II we have
\begin{eqnarray}
B_{1,2}=B_0\bigg[A_{0}\mathbf{somb}\bigg(\frac{2\pi{R}}{\lambda_2}\bigg|\frac{\vec{\rho}_1}{d_2+(\lambda_1/2
\lambda_2)L_1}\bigg|\bigg)+e^{i\varphi'} A_{\vec{a}}\mathbf{somb}\bigg(\frac{2\pi{R}}{\lambda_2}\bigg|\frac{\vec{a}}
{d'_2}+\frac{\vec{\rho}_1}{d_2+(\lambda_1/2\lambda_2)L_1}\bigg|\bigg)\bigg],\label{eq:Psi22f}
\end{eqnarray}
where $\varphi'=K_2\big[\frac{|\vec{a}|^2}{2}((\frac{1}{L_2}+\frac{1}{d'_2})-\frac{\vec{\rho}_2\cdot\vec{a}}{L_2}\big].$
The radius of the Airy disk is
\begin{eqnarray}
\xi=0.61\frac{\lambda_2}{R}\bigg(\frac{\lambda_1}{2\lambda_2}L_1+d_2\bigg).\label{eq:diskrudius2}
\end{eqnarray}
If $L_1\gg{d_2}$, $\xi\rightarrow\frac{0.61L_1}{R}(\frac{\lambda_1}{2})$, so that the width of the point-spread function shrinks to one half its value compared to the classical cases.  Applying the Rayleigh criterion to see the minimum resolvable distance between two point sources in the object. From the second term of Eq.~(\ref{eq:Psi22f}) the minimum distance turns out to be
\begin{eqnarray}
a_{\mathrm{min}}=0.61\frac{d'_2\lambda_2}{R},\label{eq:arayleigh2}
\end{eqnarray}
which only is a function of the wavelength of the non-degenerate photon; therefore, no spatial resolution improvement can be achieved compared to classical optics.

Finally, we consider the configuration shown in Fig.~\ref{fig:otherconfigurations}(b) which was analyzed in \cite{wen1} where it was shown that no well-defined images could be obtained.

It is straightforward to generalize the above two configurations with use of the $|1,N\rangle$ state. By replacing the source state by the state $|1,N\rangle$ in Fig.~\ref{fig:otherconfigurations}(a), it can be shown that the radius of the Airy disk becomes
\begin{eqnarray}
\xi=0.61\frac{\lambda_2}{R}\bigg(\frac{\lambda_1}{N\lambda_2}L_1+d_2\bigg).\label{eq:diskrudiusN}
\end{eqnarray}
If $L_1\gg{d_2}$, $\xi\rightarrow\frac{0.61L_1}{R}(\frac{\lambda_1}{N})$, so the Airy disk shrinks to one $N$th of its radius compared to classical optics. However, if $L_1\ll{d_2}$, Eq.~(\ref{eq:diskrudiusN}) gives the same result as in
classical optics. Replacing the source with photon state $|1,N\rangle$ in Fig.~\ref{fig:otherconfigurations}(b), the above conclusion is still valid. The analysis has been presented in \cite{wen2} and we will not repeat here.

\section{Conclusions}
In summary, we have proposed a quantum-imaging scheme to improve the spatial resolution in the object beyond the Rayleigh diffraction limit by using an entangled photon-number state $|1,N\rangle$. We have shown that by sending the $N$ degenerate photons to the object, keeping the non-degenerate photon and imaging lens in the lab, and using a resolving $N$-photon detector or a bucket detector, a factor of $N$ can be achieved in spatial resolution enhancement using the Rayleigh criterion. The image is nonlocal and the quantum nature of the state leads to the sub-Rayleigh imaging resolution with high contrast. We have also shown that by sending the $N$ degenerate photons freely to a point $N$-photon detector while propagating the non-degenerate photon through the imaging lens and the object, the Airy disk in the imaging can be shrunk by a factor of $N$ under certain conditions. However, it may be possible to show that a similar effect can occur using non-entangled sources. In the language of quantum information, the non-degenerate photon may be thought of as an ancilla onto which the information about the object is transferred for measurement. Our imaging protocol may be of importance in many applications such as imaging, sensors, and telescopy.

\section{Acknowledgement}
This work was supported in part by U.S. ARO MURI Grant W911NF-05-1-0197 and by Northrop Grumman Corporation through the Air Force Research Laboratory under contract FA8750-07-C-0201 as part of DARPA's Quantum Sensors Program.

\appendix
\section{Validity of the Assumption Made in Eq.~(\ref{eq:Psiassumption})}
In going from Eq.~(\ref{eq:psiapproximation}) to Eq.~(\ref{eq:Psiassumption}), we have made an assumption that
requires the detector D$_1$ is only sensitive to the scattered photons from the same spatial point in the object. This allowed us to collapse the $N$ integrations over the object into a single integral. In this Appendix, we give an example of how this assumption may be satisfied for multi-photon scattering off the target. Our example assumes that each point of the object transmits or scatters the light with a random phase which satisfies Gaussian statistics.  The result is that the visibility decreases.

We start with the case of $2+1$ photons. From Eq.~(\ref{eq:psiapproximation}) the integration over the transverse vector $\vec{\alpha}_1$, which gives
\begin{eqnarray}
B_{1,2}&=&B_0\int{d}^2\rho_oA(\vec{\rho}_o)e^{i\frac{K_1|\vec{\rho}_o|^2}{4}[\frac{2}{L_1}+\frac{1}{d_1+(2\lambda_2/
\lambda_1)d_2}]}e^{-i\frac{K_1\vec{\rho}_{1,1}\cdot\vec{\rho}_o}{L_1}}e^{i\phi(\vec{\rho}_o)}\int{d}^2\rho'_oA(\vec
{\rho}'_o)e^{i\frac{K_1|\vec{\rho}'_o|^2}{4}[\frac{2}{L_1}+\frac{1}{d_1+(2\lambda_2/\lambda_1)d_2}]}\nonumber\\
&&\times{e}^{-i\frac{K_1\vec{\rho}_{1,2}\cdot\vec{\rho}'_o}{L_1}}e^{i\phi(\vec{\rho}'_o)}e^{i\frac{K_1\vec{\rho}_o\cdot
\vec{\rho}'_o}{2[d_1+(2\lambda_2/\lambda_1)d_2]}}\int{d}^2\rho_le^{i\frac{K_2|\vec{\rho}_l|^2}{2}[\frac{1}{L_2}+\frac{1}
{d_2+(\lambda_1/2\lambda_2)d_1}-\frac{1}{f}]}e^{-iK_2\vec{\rho}_l\cdot[\frac{\vec{\rho}_2}{L_2}+\frac{\vec{\rho}_o+\vec{
\rho}'_o}{2d_2+(\lambda_1/\lambda_2)d_1}]},\label{eq:completeintegration}
\end{eqnarray}
where $\vec{\rho}_{1,j}$ is a point at which a photon is detected on the bucket detector, each point of the amplitude has a random phase associated with its transmission amplitude and, as usual, all the slowly varying terms have been grouped into $B_0$. Using the the Gaussian thin-lens imaging condition (\ref{eq:GTLE1}) gives
\begin{eqnarray}
B_{1,2}&=&B_0\int{d}^2\rho_oA(\vec{\rho}_o)e^{i\frac{K_1|\vec{\rho}_o|^2}{4}[\frac{2}{L_1}+\frac{1}{d_1+(2\lambda_2/
\lambda_1)d_2}]}e^{-i\frac{K_1\vec{\rho}_{1,1}\cdot\vec{\rho}_o}{L_1}}e^{i\phi(\vec{\rho}_o)}\int{d}^2\rho^{\prime}_o
A(\vec{\rho}'_o)e^{i\frac{K_1|\vec{\rho}^{\prime}_o|^2}{4}[\frac{2}{L_1}+\frac{1}{d_1+(2\lambda_2/\lambda_1)d_2}]}
\nonumber\\
&&\times{e}^{-i\frac{K_1\vec{\rho}_{1,2}\cdot\vec{\rho}'_o}{L_1}}e^{i\phi(\vec{\rho}'_o)}e^{i\frac{K_1\vec{\rho}_o\cdot
\vec{\rho}'_o}{2[d_1+(2\lambda_2/\lambda_1)d_2]}}\mathbf{somb}\bigg(\frac{2\pi{R}}{\lambda_2}\bigg|\frac{\vec{\rho}_2}
{L_2}+\frac{\vec{\rho}_o+\vec{\rho}^{\prime}_o}{2d_2+(\lambda_1/\lambda_2)d_1}\bigg|\bigg).\label{eq:completeintegration2}
\end{eqnarray}

Generalizing to the case of $N+1$, using the Gaussian thin-lens equation (\ref{eq:GTLEn})
\begin{eqnarray}
B_{1,N}&=&B_0\int{d}^2\rho_{o,1}A(\vec{\rho}_{o,1})e^{i\frac{K_1|\vec{\rho}_{o,1}|^2}{2L_1}}e^{-i\frac{K_1\vec{\rho}_
{1,1}\cdot\vec{\rho}_{o,1}}{L_1}}e^{i\phi(\vec{\rho}_{o,1})}\cdots\int{d}^2\rho_{o,N}A(\vec{\rho}_{o,N})e^{i\frac{K_1|
\vec{\rho}_{o,N}|^2}{2L_1}}e^{-i\frac{K_1\vec{\rho}_{1,N}\cdot\vec{\rho}_{o,N}}{L_1}}\nonumber\\
&&\times{e}^{i\phi(\vec{\rho}_{o,N})}e^{i\frac{K_1|\vec{\rho}_+|^2}{2[d_1+(\lambda_2/N\lambda_1)d_2]}}
\mathbf{somb}\bigg(\frac{2\pi{R}}{\lambda_2}\bigg|\frac{\vec{\rho}_2}{L_2}+\frac{\vec{\rho}_+}{d_2+(\lambda_1/N\lambda
_2)d_1}\bigg|\bigg),\label{eq:completeintegralN2}
\end{eqnarray}
where $\vec{\rho}_+=\frac{1}{N}\sum_{j=1}^N\vec{\rho}_{o,j}$.

To compute the counting rate we first calculate the magnitude square of the amplitude averaged over the random phases.
Starting with the $N=2$ case and assuming that the ensemble average, $\langle\cdots\rangle$, over those phases satisfies Gaussian statistics so that
\begin{eqnarray}
\langle{e}^{i[\phi(\vec{\rho}_o)+\phi(\vec{\rho}'_o)-\phi(\vec{\rho}''_o)-\phi(\vec{\rho}'''_o)]}\rangle=\delta(\vec{
\rho}_o-\vec{\rho}''_o)\delta(\vec{\rho}'_o-\vec{\rho}'''_o)+\delta(\vec{\rho}_o-\vec{\rho}'''_o)\delta(\vec{\rho}'_o
-\vec{\rho}''_o),\label{eq:deltaproperty}
\end{eqnarray}
We have assumed that the correlation length of the random phase is sufficiently small so that the Gaussian distribution can be approximated by delta functions. We find
\begin{eqnarray}
\langle{B}^*_{1,2}B_{1,2}\rangle=|B_0|^2\int{d}^2\rho_o\int{d}^2\rho'_o|A(\vec{\rho}_o)A(\vec{\rho}'_o)|^2\mathbf{somb}
^2\bigg(\frac{2\pi{R}}{\lambda_2}\bigg|\frac{\vec{\rho}_2}{L_2}+\frac{\vec{\rho}_o+\vec{\rho}'_o}{2d_2+(\lambda_1/
\lambda_2)d_1}\bigg|\bigg)\bigg[1+e^{-i\frac{K_1(\vec{\rho}_1-\vec{\rho}'_1)\cdot(\vec{\rho}_o-\vec{\rho}'_o)}{L_1}}
\bigg].\label{eq:Bsquare2}
\end{eqnarray}
When we integrate over the bucket detector, the first term will be a constant while the second term will give us a delta function in $\vec{\rho}_o$ times the area of the bucket detector, $s_b$. Equation~(\ref{eq:Bsquare2}) reduces to
\begin{eqnarray}
\int{d}^2\rho_{1,1}\int{d}^2\rho_{1,2}\langle|B_{1,2}|^2\rangle=C+|B_0|^2s_b^2(\frac{L_1\lambda_1}{2\pi{s_b}})\int{d}^2
\rho_o|A(\vec{\rho}_o)|^4\mathbf{somb}^2\bigg(\frac{2\pi{R}}{\lambda_2}\bigg|\frac{\vec{\rho}_2}{L_2}+
\frac{\vec{\rho}_o}{d_2+(\lambda_1/2\lambda_2)d_1}\bigg|\bigg).\label{eq:Bsquarepoint}
\end{eqnarray}
First note that for the second term is similar to Eq.~(\ref{eq:bucket}), the difference being the term in parenthesis which is the ratio of effect of diffraction to the area of the bucket detector, it is essentially the inverse of the Fresnel number. Computing the constant, $C$, is generally difficult and depends in detail on the geometry of the object, we can obtain an upper bound on $C$ quite easily,
\begin{equation}
|C|\leq{s}_b^2|B_0|^2\bigg|\int{d}^2\rho_o|A(\vec{\rho}_o)|^2\bigg|^2,\label{eq:bound}
\end{equation}
consequently, the visibility will be much less than for the ideal case discussed above.  From Eq.~(\ref{eq:Bsquarepoint}) the second term is proportional to $L_1 \lambda_1$ which implies that as this product increases the visibility increases, however, recall for the case of sensors $L_1 \simeq d_1$, so as this term increases the minimum resolvable distance also increases.

The generalization to the case of $N+1$ photons is straightforward. The ensemble phase average now becomes
\begin{eqnarray}
\left\langle\mathrm{exp}\bigg[i\bigg(\sum_{j=1}^N\phi(\vec{\rho}_{oj})-\sum_{j=1}^N\phi(\vec{\rho}'_{oj})\bigg)
\bigg]\right\rangle=\sum_{P_N}\prod_{r=1}^N\delta(\vec{\rho}_{o,r}-\vec{\rho}'_{o,P_N(r)}),\label{eq:deltapropertyN}
\end{eqnarray}
where the $N$ degenerate transmitted or reflected photons acquire random phases $\phi(\vec{\rho}_{o,j})$ and $P_N$ is the set of permutations of the numbers $(1,\cdots,N)$. In Eq.~(\ref{eq:deltapropertyN}) there are $N!$ terms. We can show that
\begin{eqnarray}
\langle|B_{1,N}|^2\rangle&=&|B_0|^2\int{d}^2\rho_{o,1}\cdots\int{d}^2\rho_{o,N}|A(\vec{\rho}_{o,1})\cdots{A}(\vec{\rho
}_{o,N})|^2\mathbf{somb}^2\bigg(\frac{2\pi{R}}{\lambda_2}\bigg|\frac{\vec{\rho}_2}{L_2}+\frac{\vec{\rho}_+}{N[d_2+(
\lambda_1/N\lambda_2)d_1]}\bigg|\bigg)\nonumber\\
&&\times\sum_{P_N}e^{-i\frac{K_1}{L_1}\sum_{r=1}^N{\vec{\rho}_{1,r}\cdot(\vec{\rho}_{o,r}-\vec{\rho}_{o,P_N(r)})}}.
\label{eq:BsquareN}
\end{eqnarray}
When we integrate over the bucket detector, we get a complicated result. Two terms are simple, the identity permutation gives a constant and the single cycle subgroup give an incoherent image with a resolution that depends on $\lambda_1/N$. These are the only terms for $N=2$. The remaining terms will lead to terms which are essentially constant. For $N=3$ we get
\begin{eqnarray}
\lefteqn{\int{d}^2\rho_{1,1}\int{d}^2\rho_{1,2}\int{d}^2\rho_{1,3}\langle|B_{1,3}|^2\rangle=C+3|B_0|^2s^3_b\bigg(\frac{
L_1\lambda_1}{2\pi{s}_b}\bigg)\int{d}^2\rho_{+}\int{d}^2\zeta|A(\vec{\rho}_{+}-2\vec{\zeta})|^2 |A(\vec{\rho}_{+}+\zeta)|^4}\nonumber\\
&&\times\mathbf{somb}\bigg(\frac{2\pi{R}}{\lambda_2}\bigg|\frac{\vec{\rho}_2}{L_2}+\frac{\vec{\rho}_{+}}
{3[d_2+(\lambda_1/3\lambda_2)d_1]}\bigg|\bigg)+|B_0|^2s^3_b\bigg(\frac{L_1\lambda_1}{2\pi{s}_b}\bigg)^2\int{d}
^2\rho_{o,1}|A(\vec{\rho}_{o,1})|^6\nonumber\\
&&\times\mathbf{somb}^2\bigg(\frac{2\pi{R}}{\lambda_2}\bigg|\frac{\vec{\rho}_2}{L_2}+\frac{\vec{\rho}_{o,1}}{d_2+(\lambda
_1/N\lambda_2)d_1}\bigg|\bigg).\label{eq:Bsquarebucket}
\end{eqnarray}
From Eq.~(\ref{eq:Bsquarebucket}) the second term shows explicitly how the general terms will lead to a complicated average over the illuminated area of the object. This result shows that the image will have very poor visibility for large $N$, it is not certain whether there might be arrangement of detectors for the $N$ photons which will give better results.


\begin{thebibliography}{99}
\bibitem{rayleigh} It should be noted that there are different, related, meanings of the Rayleigh criterion. The one used here refers to the ability to resolve two point sources in the object. The other is the minimum angle between plane waves falling on an aperture that can be resolved and can be interpreted as resolving two point sources at infinity.
\end{thebibliography}
\end{document}